%% file: ms.tex
\documentclass[12pt,preprint]{aastex52/aastex}

\slugcomment{submitted to the ApJL}

\newcommand       \mic        	 {$\mu$m}
\newcommand       \nwat        	 {nW~m$^{-2}$~sr$^{-1}$}
\newcommand       \spitz        	 {{\it Spitzer}}
\newcommand       \hst        	 {{\it HST}}
\newcommand       \cobe        	 {{\it COBE}}

\begin{document}

\def\lsim{\mathrel{\hbox{\rlap{\hbox{\lower4pt\hbox{$\sim$}}}\hbox{$<$}}}}
\def\gsim{\mathrel{\hbox{\rlap{\hbox{\lower4pt\hbox{$\sim$}}}\hbox{$>$}}}}

 \title{Constraints to Energy Spectra of Blazars based on  Recent EBL Limits from Galaxy Counts }

\author{F. Krennrich\altaffilmark{1}, E. Dwek\altaffilmark{2} \& A. Imran\altaffilmark{1}   }

\altaffiltext{1}{Department of Physics and Astronomy, Iowa State University, Ames, IA, 5011-3160, USA,
e-mail: krennrich@iastate.edu}
\altaffiltext{2}{Observational Cosmology Laboratory, Code 665, NASA Goddard Space Flight Center, Greenbelt, MD 20771, e-mail: eli.dwek@nasa.gov}

\begin{abstract}

We combine the recent estimate of the contribution of galaxies to the 3.6~\mic\ intensity of the extragalactic background light (EBL) with optical and near-infrared (IR) galaxy counts to set new limits on intrinsic spectra of some of the most distant TeV blazars 1ES~0229+200, 1ES~1218+30.4, and 1ES~1101-232, located at redshifts 0.1396, 0.182, and 0.186, respectively.  
The new lower limit on the 3.6~\mic\ EBL intensity is significantly higher than the previous one set by the cumulative emission from resolved \spitz\ galaxies. Correcting for attenuation by the revised EBL, we show that the differential spectral index of the intrinsic spectrum of the three blazars is $\rm 1.28 \pm 0.20$  or harder.  These results present blazar emission models with the challenge of producing extremely hard intrinsic spectra in the sub-TeV to multi-TeV regime.
These results also question the reliability of recently derived upper limits on the near-IR EBL intensity that are solely based on the assumption that intrinsic blazar spectra should not be harder than 1.50 (Aharonian et al. 2006; Aharonian et al. 2007a).

\end{abstract}


\section{INTRODUCTION}

The absorption of TeV photons on intergalactic distance scales via $\rm  \gamma \gamma \rightarrow  e^{+} \: e^{-} $ by the diffuse radiation from the extragalactic background light (EBL) is crucial for the understanding of  TeV $\gamma$-ray sources with significant redshift (Gould \& Schr\'eder 1967). Direct measurements of the EBL at near- to mid-IR wavelengths are greatly hampered by the presence of strong foreground emission from interplanetary dust and diffuse Galactic emission from unresolved stars and the interstellar medium. Consequently, our knowledge of the EBL consists mostly of lower limits derived from ground- and space-based galaxy counts or direct measurements  prone to substantial systematic uncertainties (Hauser \& Dwek 2001).


The catalog of TeV blazars now contains more than 20 objects (see, e.g., Hinton 2007) spanning a large range of redshifts between 0.03 and 0.536. Their observed energy spectra have therefore passed through different path lengths resulting in different degrees of attenuation. The observed TeV energy spectra of blazars  therefore contain information about their intrinsic spectra convolved with the absorption due to the EBL in the near-IR and mid-IR. Separating the intrinsic blazar spectra from the absorption effect to derive constraints to the EBL  has proven a difficult task (see Stecker et al. 1992; Dwek \& Krennrich 2005; Aharonian et al. 2006; Stecker et al. 2007;  Mazin \& Raue 2007)  and may require a significantly larger sample of TeV blazars. However, if successful,  the potential of providing constraints on the EBL density  using TeV energy spectra provides complementary  information to direct measurements of the EBL (see, e.g.,  Coppi et al. 1999).

Galaxy counts provide a firm lower limit on the EBL, and therefore a strong lower limit on the magnitude of absorption  present in a blazar spectrum.  Figure 1 presents 0.36, 0.45, 0.67, 0.82, 1.1, 1.6 and 2.2~\mic\ lower limits on the EBL determined from ground-based and {\it Hubble Space Telescope} (\hst) observations (filled triangles) presented  by Madau \& Pozetti (2000) (hereafter, MP00). At 3.6, 4.5, 5.8, and 8.0~\mic\ (open quadrangles), lower limits arise from galaxy counts using the IRAC instrument on the \spitz\ infrared space observatory (Fazio et al. 2004), and the 15 and 24~\mic\ \spitz/MIPS data, obtained by Metcalfe et al. (2003) and Papovich et al. (2004),  respectively (hollow circle and diamond). Recently, Levenson \& Wright (2008) (hereafter, LW08) used profile fitting techniques to estimate the contribution of the faint fuzzy fringes of galaxies to the total contribution of resolved galaxies, increasing the 3.6~\mic\ lower limit from 5.4~\nwat\ to 9.0$^{+1.7}_{-0.9}$~\nwat. This correction has brought the galaxies' contribution to the EBL to within $\sim 1\sigma$ of the EBL intensity determined from measurements made with the {\it Cosmic Background Explorer} (\cobe) satellite (Dwek \& Arendt 1998). In the following we treat the constraint by LW08 purely as a lower limit to the EBL at 3.6~micron, consistent with their interpretation of the result.

The same considerations that have resulted in a significant underestimate of the contribution of resolved galaxies to the 3.6~\mic\ EBL may also apply to the estimated integrated light from resolved galaxies obtained from standard aperture photometry (Bernstein et al. 2002; Wright 2001; Yoshi  1993; Bernstein 2007). The galaxies' contribution to the EBL could therefore be higher than the reported lower limits of MP00. The combination  of the new limit at 3.6~micron with potentially higher limits at shorter wavelengths (MP2.0+LW+MIR) will imply  more attenuation for $\gamma$-ray spectra at TeV energies and  harder intrinsic spectra. 
In this paper we describe results from exploring a wide range of possible intrinsic energy spectra  compatible with EBL lower limits. 


Only a limited number of blazar spectra at TeV energies at  sufficiently high redshift, and consequently substantial  absorption are available.  This lead us to select the energy spectra of  1ES~1218+30.4 (Albert et al. 2006) detected over the 0.08 to 0.7~TeV range, and also recently detected by VERITAS over an energy range of 0.18 to 1.5~TeV (Fortin et al. 2008)\footnote{We use the VERITAS spectrum as it extends to higher energies than that of MAGIC.}; 1ES1101-232 (Aharonian et al. 2006) detected over the 0.18 to 2.9~TeV range, and  1ES~0229+200 (Aharonian et al. 2007a) detected over the 0.6 to 11.5~TeV range.
Note that 3C~279, the most distant blazar at a redshift of 0.536, has only been observed over a very limited spectral range from 80 to 485~GeV, and has considerably larger statistical errors than those presented above (Albert et al. 2008).   Other blazars with adequately measured energy spectra are 1ES~1011+496 (Albert et al. 2007) and 1ES~0347-121 (Aharonian et al. 2007b) with redshifts of z=0.212 and z=0.188, respectively. The former is not relevant in the search for a limit to the hardness of blazar spectra since its measured energy spectrum is extremely soft, with $\Gamma = 4.0 \pm 0.5$.  The BL Lacertae 1ES~0347-121 has a spectrum that is slightly softer ($\Gamma = 3.10   \pm 0.23$ ) than that of 1ES~1101-232 and covers essentially the same energy range and therefore is slightly less constraining and  redundant.  We therefore limit our selection of blazars primarily to  1ES~1218+30.4, 1ES~1101-232 and 1ES~0229+200. They constitute a representative sample of the most distant, well measured blazar spectra, covering a wide energy range with data provided by three independent $\gamma$-ray  observatories (MAGIC, HESS \& VERITAS).    Because the cross section for the $\gamma-\gamma$ interaction peaks at energies $E_{\gamma}$(TeV)$\approx 0.8 \lambda$(\mic), the intrinsic spectrum of each blazar will be constrained by different spectral regions of the EBL.  The set of these three blazars  therefore allows us to determine the intrinsic hardness of blazar spectra over a wide range of energies.

The  aim of this study is to derive the range of possible spectral indices that characterize the intrinsic energy spectra of blazars.   This allows one to use observations to test the previously postulated theoretical paradigm, that the intrinsic spectra of blazars characterized by an energy spectrum of the form $\rm  dN/dE \propto E^{-\Gamma_{i}}$  in the TeV regime (Aharonian et al. 2006)  cannot exceed the hardness of $\Gamma_{i}= 1.5$.


\section{THE INTRINSIC BLAZAR SPECTRA FOR DIFFERENT EBL SCENARIOS}

In order to obtain the intrinsic spectrum of a blazar, the measured spectrum must be corrected for the effects of the EBL. Figure~1 shows a wide range of EBL intensities (shaded area)  considered for unfolding the intrinsic spectra.    The lower limits from galaxy counts restrict the possible EBL density as a function of wavelength.  A convenient parameterization of EBL scenarios  and a wide range of EBL spectra with different near-IR to mid-IR ratios was provided by Dwek \& Krennrich (2005) and are used in this study for deriving a limit to the hardness of the blazar spectra.
For illustrative purposes we use a low EBL scenario that is  called LLL for Low near-IR, Low mid-IR, Low far-IR which is consistent with the limits from MP00 but falls significantly  below the galaxy counts in the mid-IR.  Furthermore, we have considered a  variety of EBL scenarios  with different spectral indices and shapes that are within the boundaries of the shaded area in Figure~1.  The upper bound is somewhat arbitrary, however, as will become apparent, uniformly higher density EBL scenarios lead to even harder intrinsic spectra and are moot.  The shaded area also covers the EBL wavelength regime that is most sensitive to the blazar energy spectra in question, nevertheless, for completeness we also include the limits at 15~$\mu$m and at 24~$\mu$m.

  We also show results for an EBL spectrum that was presented in Aharonian et al. (2006) and is dubbed AHA0.65 (P0.65 in original paper).  In addition, we widen our range of EBL scenarios allowing for a scaling factor in $\rm \nu I_{\nu}$ which follows the name of the EBL scenario.   Finally, we also consider effects on the intrinsic energy spectra imposed by lower limits at 15~$\mu$m (Metcalfe et al. 2003) and 24~$\mu$m (Papovich et al. 2004).  Those scenarios carry the suffix 'MIR'.
  Also note that Figure 1 shows for sake of clarity only a limited set of the explored scenarios, the ones that are most relevant for the subsequent discussion. Figure 1 also illustrates our principal approach in exploring different spectral slopes between the near-IR and the mid-IR: the MP+LW+MIR and MP2.0+LW+MIR scenarios show two different spectral slopes while both spectra intersect with the lower limit at 3.6~$\mu$m. We also show a scenario MP+0.7LW+MIR  which falls substantially below the limit at 3.6~$\mu$m.
Two different EBL codes were used to derive the intrinsic spectra, which allowed us to test and verify the results independently.  The derived spectral indices characterizing the intrinsic blazar spectra for the different EBL scenarios are listed in Table 1.



The three selected blazars constrain the EBL in different wavelength regimes, so that the combination of the three, simultaneously probing different EBL scenarios in the  near-IR to  mid-IR, provides the strongest constraint to their intrinsic spectra.  The methodology to find a limit to the intrinsic spectra of these three by applying EBL scenarios goes as follows: we look for  the hardest spectral index that any of the three blazars show for a given EBL scenario and then search for the EBL scenario that allows for  the softest spectrum among those.

The results are presented in Table~1  showing the spectral indices $\rm \Gamma_{i}$ for power law fits to the absorption corrected (also referred to as intrinsic) energy spectra of 1ES~1101-232, 1ES~1218+30.4 and 1ES~0229+200 for a range of EBL scenarios.  To give the reader an idea about the dependence of spectral index on EBL density, we also show scenarios that fall below limits from galaxy counts, e.g., the LLL scenario. Those  are marked with an asterisk in Table~1 and are not viable, e.g., the LHL0.76, the LLL and the LLH scenarios, etc.
The ones that are still compatible with the lower limits from galaxy counts are shown in the two lower sections  of Table~1 (separated by double line) and  the absorption corrected $\gamma$-ray spectra collectively give a lower limit to the hardness of these blazar spectra. We furthermore show three additional scenarios (below single line in Table 1) that take into account lower EBL limits from galaxy counts at 15~$\mu$m and 24~$\mu$m: AHA0.65+MIR constitutes the standard AHA0.65 scenarios up to 10~$\mu$m plus an MIR component consistent with lower limits.  The MP+LW+MIR scenario is based on a fit through  the MP00 data, the LW08 data point and the MIR data. The MP2.0+LW+MIR differs from the last case by its level in the near-IR with the   MP00 values scaled up by a factor of two.

For example, when just considering 1ES~1218+30.4 by itself, its softest intrinsic spectrum among all EBL scenarios that obey the galaxy count limits, is described by a power law with index $\rm \Gamma_{i} = 1.83 \pm 0.40$ and corresponds to the LHL1.25 scenario.  However, for the same EBL scenario, the spectrum of 1ES~0229+200 would be extremely hard  showing a power law index of $\rm \Gamma_{i} = 0.59 \pm 0.34$.  By searching for the combination of intrinsic spectra of these three sources that yield the softest spectral indices, one finds that the AHA0.65 scenario gives a spectral index $\rm \Gamma_{i} = 1.28 \pm 0.20$ for 1ES~1101-232, $\rm \Gamma_{i} = 1.30 \pm 0.38$ for 1ES~1218+30.4,   while 1ES~0229+200 gives  $\rm \Gamma_{i} =  2.40 \pm 0.13$.  The spectral index of $\rm  \Gamma_{i} = 1.28 \pm 0.20$ corresponds to the least hard spectral index of all EBL scenarios (shaded region) consistent with the lower limits.
In other words, any other EBL scenario compatible with the lower limits from galaxy counts results in one of the three blazar spectra to be harder than  $\rm \Gamma_{i} = 1.28 \pm 0.20$.  The lower limit from LW08 is an  important new constraint to the intrinsic spectra;  the same analysis carried out with the lower limits provided by Fazio et al. (2004) lead to a spectral index of $\rm \Gamma_{i} = 1.78 \pm 0.20$ making it significantly less constraining than the recently improved lower limit.

The addition of an MIR component consistent with the  15~$\mu$m and 24~$\mu$m constraints does not change the result significantly.  Finally, we show the absorption corrected (AHA0.65) energy spectra of  1ES~1218+30.4, 1ES~1101-232 and 1ES0229+200  in Figure~2. The former two appear to have their peak energy above 2~TeV.  Since the EBL used for correcting for absorption is the AHA0.65 scenario, which is somewhat above the lower limits from HST galaxy counts, it is entirely possible that the source spectra of those two blazars are softer, however, in that case the source spectrum of 1ES~0229+200 would have to be significantly harder, again bringing the peak energy of 1ES~0229+200 well above 10~TeV.  Also note the presence of a bump in the corrected spectrum of 1ES~0229+200, although the probability for a power law fit is 5.5\%. The sharpness of this feature indicates that this feature is likely due to an over correction for EBL absorption.
However,  neither the AHA0.65+MIR, nor the MP+LW+MIR, the MP2.0+LW+MIR and the MP+0.7LW+MIR show  a significant bump, suggesting that this feature is only an artifact that disappears when the EBL is extended to include the mid-IR limits.

Our results for the intrinsic energy spectra of blazars is entirely based on observational data and therefore provides a firm limit to the hardness of the energy spectra of some of the most distant TeV blazars.  The simultaneous application of EBL scenarios to
 blazar spectra that provide sensitivity to two different EBL wavelength regimes, the 0.2 - 3.6~$\mu$m (1ES~1101-232, 1ES~1218+30.4) and the 0.7 - 14~$\mu$m (1ES~0229+200)\footnote{The power law indices of the blazar spectra of 1ES~1101-232, 1ES~1218+30.4 are primarily dominated by $\gamma$-ray energies of 0.18 to 1 TeV considering the statistical uncertainties making
these spectra mostly sensitive to 0.2 - 1.2~$\mu$m while the spectral index of 1ES~0229+200 is dominantly determined
by photon energies between 1 - 4 TeV making it mostly sensitive to 1.2 - 5~$\mu$m; the spectral coverage at $\gamma$-ray energies translates into complementary spectral constraints in the EBL waveband.} combined with the new lower limit at 3.6~$\mu$m leads to much stronger and unambiguous constraints to the intrinsic spectra than any previous study.

\section{Conclusions}

In this paper we have presented observational evidence that the absorption corrected energy spectra of individual TeV blazars are extremely hard, exhibiting spectra with an index of   $\rm \Gamma_{i} = 1.28 \pm 0.20$ or harder.  In fact the spectra are likely even harder considering that the lower limits from MP00 do not take into account the faint and fuzzy fringes of resolved galaxies. If  their lower limits will be scaled up by a factor of 2, then MP2.0+LW+MIR may be a more realistic EBL scenario,  a slightly harder intrinsic spectrum for 1ES~1101-232 with an index of   $\rm \Gamma_{i} = 1.18 \pm 0.20$.  One should also note that this  EBL scenario is similar to the EBL model from Primack (2005) used in Aharonian et al. (2006) to set upper limits to the EBL. The results of our studies show that the range of the intrinsic spectral indices of the three blazars is in the range of $\Gamma_{i} = 1.18 \pm 0.20$, and no softer than $1.29 \pm 0.20$ (AHA0.65+MIR). These results are a direct consequence of the new limit on the EBL, and pose a significant challenge to models that suggest that intrinsic blazar spectra cannot be harder than $\Gamma_{i} = 1.5$.  This result concurs well with a recent paper by Stecker et al. (2007) in which the authors show that the energy spectral index of the blazar 1ES~0229+200 could have  a hard intrinsic spectral index between $\rm 1.1 \pm 0.3 $ and $\rm 1.5 \pm 0.3$ if their EBL models are correct.   We confirm the hardness of the energy spectral indices based on a new observational constraint and a new analysis method. 

Franceschini et al. (2008) provided a very detailed  compilation of the EBL based on a vast amount of survey data and a backward evolution model. They applied their model EBL that includes redshift evolution to the blazar data from 1ES~1101-232 yielding a spectral index of  $\Gamma_{i} = 1.6$.  However, the EBL model did not consider the recent limit from LW08. 
As a consequence, our result of $\Gamma_{i} = 1.28 \pm 0.20 $ for the intrinsic spectra of the three blazars and their result are not conflicting, since the former is based on theoretical modeling and our result is based on observational constraints. 

The notion of potentially hard intrinsic TeV blazar spectra and their implications for EBL constraints and relativistic jets has been extensively discussed in the literature with emphasis on theoretical models (Aharonian et al. 2006; Malkov \& O'C Drury 2001; Stecker et al. 2007; Katarzy\'nski et al. 2006;  B\"ottcher et al. 2008; Aharonian 2008).   Gamma-ray spectra of blazars with  $\Gamma_{i} \le 1.5 $ were considered inconsistent with TeV spectra that originate from processes that involve  diffusive shock acceleration (Aharonian et al. 2006; Malkov \& O'C Drury 2001).   Recent numerical simulations performed by Stecker et al. (2007) however seem to indicate that sufficiently  hard electron spectra could be generated by diffuse shock acceleration at relativistic shocks. 
Even for a hard spectrum electron population,   it is argued by B\"ottcher et al. (2008) that in a Synchrotron-Self-Compton (SSC) scenario the resulting GeV - TeV $\gamma$-ray spectra would experience substantial softening  from Klein-Nishina effects making $\Gamma_{i} \le 1.5$ difficult to model.   Another solution to the problem  was proposed by Katarzy\'nski et al. (2006);  a high low-energy cutoff in the electron distribution could give the appearance of a hard $\gamma$-ray spectrum for a given energy regime. Aharonian al. (2008) show that $\rm \gamma\gamma$ absorption in the source due to narrow band emission from the AGN could lead to unusually hard  TeV spectra from AGNs.

Another interesting and potentially verifiable solution was given by B\"ottcher et al. (2008). They introduce an additional radiation component that arises from Compton up scattering of ambient photons from the Cosmic Microwave Background, occurring in a kiloparsec scale jet.  In this case, a substantial fraction of the jet power is transported by hadrons to the outer regions of the jet, where they are dissipated into ultrarelativistic electrons. Such a hard radiation component would be associated with the large scale of  the kpc jet without exhibiting the time variation that are typically seen in TeV emission from blazars.   Observations of 1ES~1101-232 (Aharonian et al. 2006), 1ES~1218+30.4 (Albert et al. 2006; Fortin et al. 2008) and 1ES~0229+200 (Aharonian et al. 2007a) in fact do not show flux variations, not inconsistent with this scenario. However, the flux levels are only a few \% of the Crab making the detection of flares at that level difficult.   More observations are required  and could provide a means of searching for a  hadronic component in TeV blazars.

The observational limit presented in this paper puts the notion of the possibility of extremely hard TeV blazar spectra on substantially more solid footing as we provide an  observational  limit of  $\Gamma_{i} \le 1.28 \pm 0.20 $ to the intrinsic spectra of a sample of three TeV blazars.  As this limit relies on the validity of recent lower limits of galaxy counts and the assumption of EBL absorption taking place in intergalactic space, it constitutes indirect evidence for extremely hard spectra of distant TeV blazars.



\subsection*{Acknowledgments}

This research is supported by grants from the U.S. Department of Energy. The authors would like to thank Wystan Benbow and Werner Hofmann for providing the spectral data for 1ES~1101-232 and 1ES~0229+200.  The authors  thank Pascal Fortin and the VERITAS collaboration for providing the data for the blazar 1ES~1218+30.4. E.D. acknowledges the support of NASA LTSA 03-0000-065. 


\clearpage

\begin{figure}
\begin{center}
\includegraphics [width=0.70\textwidth]{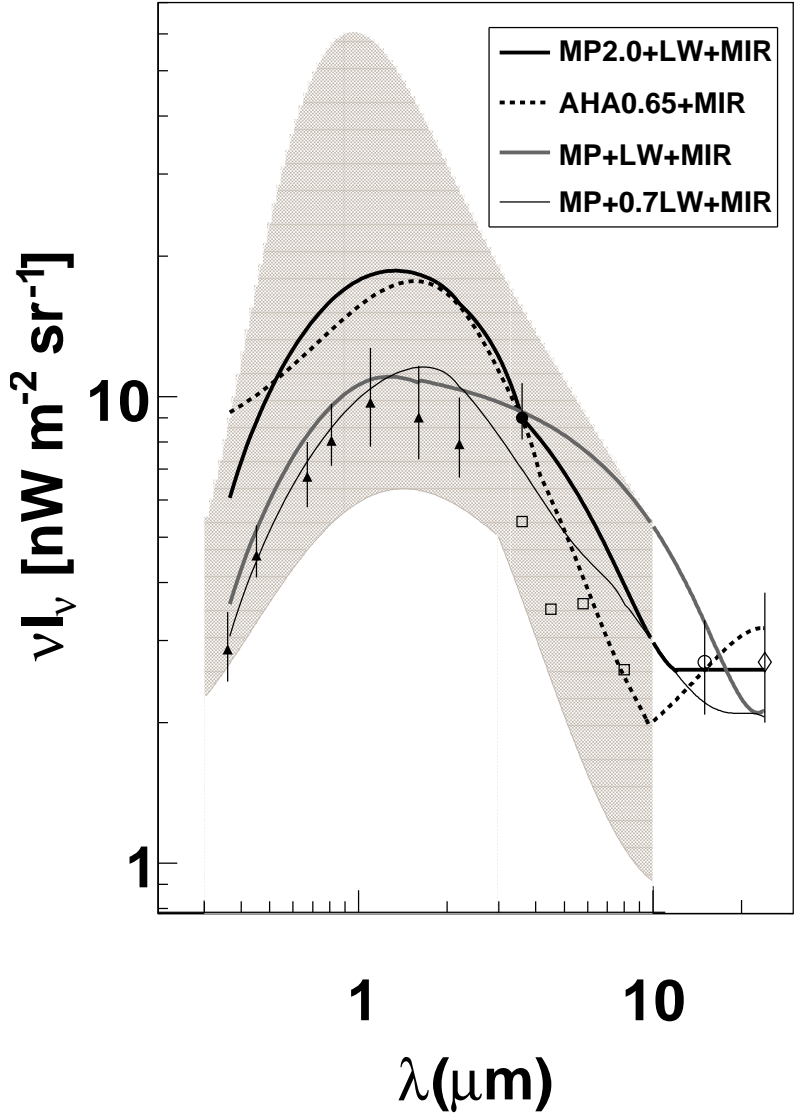}
\end{center}
\caption{Shown are select EBL scenarios to find the softest possible intrinsic blazar spectrum constrained by the lower limits from galaxy counts by Madau \& Pozetti (2000) in the optical to near-IR (solid triangles), the Spitzer results (empty squares) in the mid-IR by Fazio et al. (2004), the lower limit at 3.6$\mu$m  by LW08, a limit at 15$\mu$m by Metcalfe et al. (2003) and one at 24$\mu$m Papovich et al. (2004). The AHA0.65+MIR is motivated by the EBL model from Primack (2005) and the work presented in Aharonian et al. (2006), the MP+LW+MIR is motivated by strictly considering the galaxy counts, and finally the MP2.0+LW+MIR is motivated by the fact that the galaxy counts in the optical to near-IR could also be underestimated because of the effects described by LW08.  Furthermore, we have explored a
wide range (shaded region) of EBL scenarios from Dwek \& Krennrich et al. (2005), results are shown in Table~1.   }\label{fig1}
\end{figure}

\clearpage

\begin{figure}
\begin{center}
\includegraphics [width=1.0\textwidth]{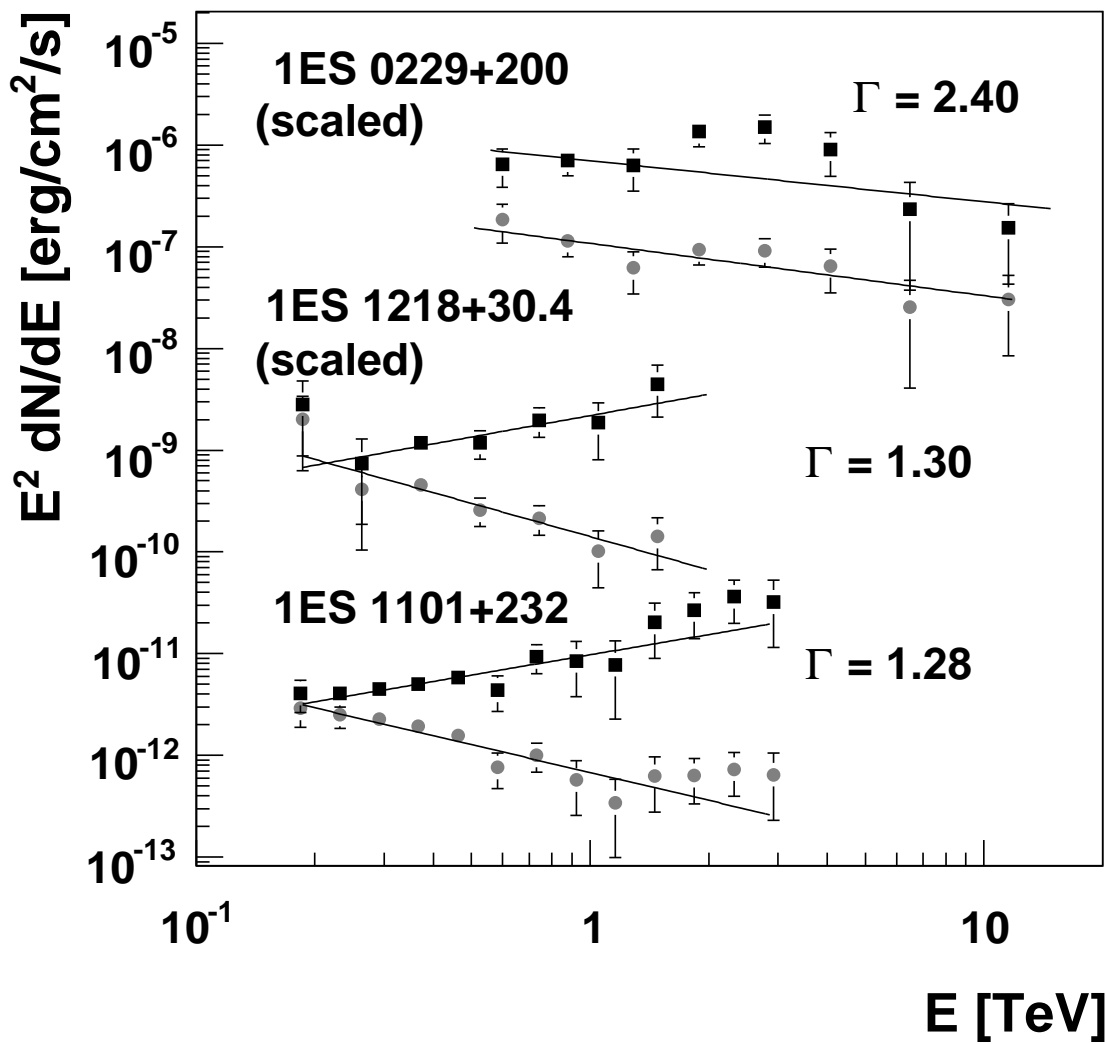}
\end{center}
\caption{Shown are the measured (filled circles) and absorption corrected (filled squares) energy spectra of 1ES~1101-232 (Aharonian et al. 2006),  1ES~1218+30.4 (Fortin et al. 2008) and 1ES~0229+200 (Aharonian et al. 2007).  Note that the latter two have been scaled for clarity. The absorption corrected spectra (solid squares) are given for the AHA0.65 scenario.  The intrinsic spectra of 1ES~1101+232 and 1ES~1218+30.4 are extremely hard, with $\rm \Gamma_{i}$ smaller than the limiting value of 1.5.}\label{fig2}
\end{figure}





\clearpage


\input{tab1.tex}

\clearpage

\end{document}

%% file: tab1.tex
\begin{deluxetable}{ccccccc}
\footnotesize
\tablecaption{Source spectra of 1ES~1101-232 and 1ES~1218+30.4 and 1ES~0229+200. Note that many scenarios  contain a suffix that represents a  scaling factor in $\rm \nu I_{\nu}$.  Scenarios marked with an asterisk are inconsistent with lower EBL limits. 
\label{tbl-1}}
\tablewidth{0pt}
\tablehead{
\colhead{Scenario} &  \colhead{} & \colhead{$\Gamma_{intrinsic}$} &
\colhead{}  & \colhead{    } \nl  \hline
\colhead{} & \colhead{1ES1101 } & \colhead{1ES1218 } &\colhead{ 1ES0229 }  }
\startdata
  AHA0.45$^*$      &   $\rm 1.78 \pm 0.20  $  &  $ \rm  1.86 \pm 0.37   $   & $ \rm  2.43 \pm 0.13   $   \nl
  LLH$^*$       &   $\rm 2.01 \pm 0.22  $  &  $ \rm  2.07 \pm 0.35   $  & $ \rm  2.12 \pm 0.20   $     \nl 
  LHL$^*$       &   $\rm 2.04 \pm 0.20  $  &  $ \rm  2.08 \pm 0.39   $  & $ \rm  0.94 \pm 0.32 $       \nl
  LHL0.76$^*$   &   $\rm 2.23 \pm 0.21  $  &  $ \rm  2.32 \pm 0.37   $  & $ \rm  1.30 \pm 0.29   $      \nl
  MHL0.70$^*$  &   $\rm 1.26 \pm 0.19  $  &  $ \rm  1.34 \pm 0.36   $  &   $ \rm  1.35 \pm 0.21  $     \nl
  MP+0.7LW+MIR$^*$ & $\rm 1.80 \pm 0.21  $  &  $ \rm  1.82 \pm 0.38   $    &  $ \rm  1.43 \pm 0.16  $     \nl 
  LLL$^*$       &   $\rm 2.06 \pm 0.16  $  &  $ \rm  2.20 \pm 0.34   $    &  $ \rm  2.11 \pm 0.20  $     \nl   \hline \hline
  HHH       &   $\rm -0.67 \pm 0.12  $  &  $ \rm  -0.72 \pm 0.29   $ & $ \rm  0.90 \pm 0.17   $     \nl
  LLL2.4    &   $\rm 1.10 \pm 0.17$  &  $ \rm  1.12 \pm 0.34 $  &  $ \rm  1.67 \pm 0.19 $     \nl
  MHL1.10    &   $\rm 0.40 \pm 0.20$  &  $ \rm  0.33 \pm 0.36 $  &  $ \rm  0.70 \pm 0.19 $      \nl
  AHA0.65    &   $\rm 1.28 \pm 0.20$  &  $ \rm  1.30 \pm 0.38 $  &  $ \rm  2.40 \pm 0.13 $      \nl
  LHL1.25    &   $\rm 1.85 \pm 0.24$  &  $ \rm  1.83 \pm 0.40 $  &  $ \rm  0.59 \pm 0.34 $     \nl  \hline 
  AHA0.65+MIR    &   $\rm 1.29 \pm 0.20$  &  $ \rm  1.31 \pm 0.38 $  &  $ \rm  1.70 \pm 0.15 $     \nl 
  MP+LW+MIR   &   $\rm 1.87 \pm 0.22$  &  $ \rm  1.77 \pm 0.42 $  &  $ \rm  1.01 \pm 0.17 $     \nl
  MP2.0+LW+MIR   &   $\rm 1.18 \pm 0.20$  &  $ \rm  1.20 \pm 0.38 $  &  $ \rm  1.53 \pm 0.15 $     \nl
   \enddata
\end{deluxetable}